\documentclass[showpacs,amsmath,amssymb,aps,prb,10pt]{revtex4}

\usepackage{graphicx}
\usepackage{dcolumn}
\usepackage{bm}

\begin{document}

\title{Spectrum of short-wavelength magnons in two-dimensional quantum Heisenberg antiferromagnet on a square lattice: third order expansion in $1/S$}

\author{A. V. Syromyatnikov}
 \email{syromyat@thd.pnpi.spb.ru}
\affiliation{Petersburg Nuclear Physics Institute, Gatchina, St.~Petersburg 188300, Russia}
\affiliation{Department of Physics, St.~Petersburg State University, Oulianovskaya 1, Petrodvorets, St.~Petersburg 198504, Russia}


\begin{abstract}

The spectrum of short-wavelength magnons in two-dimensional quantum Heisenberg antiferromagnet on a square lattice is calculated in the third order in $1/S$ expansion. It is shown that $1/S$ series for $S=1/2$ converges fast in the whole Brillouin zone except for the neighborhood of the point ${\bf k}=(\pi,0)$, at which absolute values of the third and the second order $1/S$-corrections are approximately equal to each other. It is shown that the third order corrections make deeper the roton-like local minimum at ${\bf k}=(\pi,0)$ improving the agreement with the recent experiments and numerical results in the neighborhood of this point. It is suggested that $1/S$ series converges slowly near ${\bf k}=(\pi,0)$ also for $S=1$ although the spectrum renormalization would be small in this case due to very small values of high-order $1/S$ corrections.

\end{abstract}


\maketitle

\section{Introduction}

Spin-$\frac12$ two-dimensional (2D) Heisenberg antiferromagnet (AF) on a square lattice has been one of the most attractive theoretical objects in the last two decades because this model
describes parent compounds of high-$T_c$ superconducting cuprates \cite{monous}. A number of theoretical approaches have been proposed to describe the spectrum of long-wavelength elementary excitations (magnons) in quantum square 2D AF which results agree well with each other and describe quantitatively existing experimental data \cite{monous,chak,christ}. Meantime there are some surprising recent experimental and numerical findings indicating that the standard theoretical approaches do not work for short-wavelength magnons for $S\sim1$. 

\begin{figure}
\centering
\includegraphics[scale=1.0]{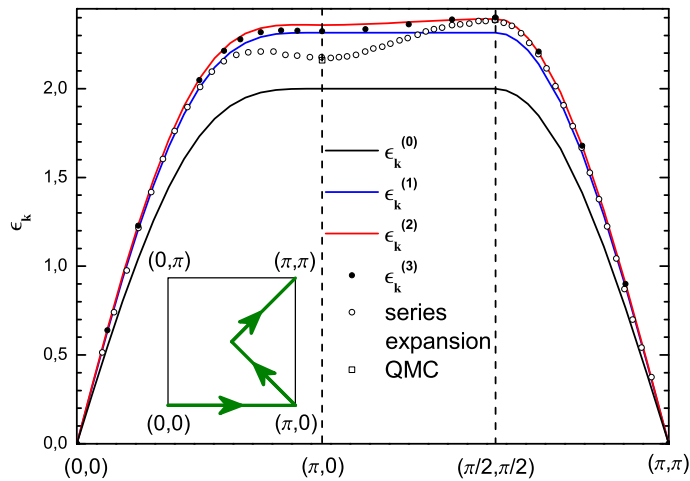}
\caption{(Color online.) Spin-wave spectrum of spin-$\frac12$ AF along high-symmetry paths of the Brillouin zone shown in the inset. Here $\epsilon_{\bf k}^{(i)}$ indicate the spectrum calculated within $i$-th order in $1/S$ so that $i=0$ corresponds to the classical spectrum \eqref{spec0}. Results of the series expansion around the Ising limit and Quantum Monte-Carlo (QMC) computation (available only for ${\bf k}=(\pi,0)$ and $(\pi/2,\pi/2)$) are also shown which were taken from Ref.~\cite{series} and Ref.~\cite{mc}, respectively. The former results describe quantitatively the spectrum observed experimentally \cite{christ} in $\rm Cu(DCOO)_2\cdot4D_2O$.
\label{spec}} 
\end{figure}

Thus, a rotonlike local minimum was observed at small $T$ in the spin-wave spectrum $\epsilon_{\bf k}$ at ${\bf k}=(\pi,0)$ in a number of recent experiments on square spin-$\frac12$ 2D AFs \cite{roton1,roton2,christ,lumsden}. In particular, the magnon energy at ${\bf k}=(\pi,0)$ appears to be 7(1)\% smaller than that at ${\bf k}=(\pi/2,\pi/2)$ in $\rm Cu(DCOO)_2\cdot4D_2O$. \cite{christ} This local minimum is purely quantum effect as the classical spectrum of 2D AF is flat along the magnetic Brillouin zone (BZ) boundary connecting points $(\pi,0)$ and $(0,\pi)$ (see inset in Fig.~\ref{spec}). The spectrum near the point ${\bf k}=(\pi,0)$ is not reproduced quantitatively within the second order in $1/S$ expansion \cite{igarashi,igar2} and phase flux RVB techniques \cite{flux}. In the former case the second order corrections lead to a very small difference of 1.4\% between $\epsilon_{(\pi,0)}^{(2)}\approx2.35858$ and $\epsilon_{(\pi/2,\pi/2)}^{(2)}\approx2.39199$ whereas in the last case this difference is too large. At the same time numerical computations using series expansion around the Ising limit \cite{series} and Quantum Monte-Carlo \cite{mc} describe the roton-like minimum satisfactorily leading to values $\epsilon_{(\pi,0)}^{(series)}\approx2.18(1)$, $\epsilon_{(\pi/2,\pi/2)}^{(series)}\approx2.385(1)$ and $\epsilon_{(\pi,0)}^{(MC)}\approx2.16$, $\epsilon_{(\pi/2,\pi/2)}^{(MC)}\approx2.39$, respectively. The origin of the local minimum has not been clarified yet. It is considered to be a signature of the spins entanglement on neighboring sites \cite{christ}. 

Existence of such a strong deviation of the spectrum near ${\bf k}=(\pi,0)$ from the result obtained in the second order in $1/S$ is quite surprising because the second order corrections are much smaller than the first order ones in the whole BZ even for $S=1/2$ (see Ref.~\cite{igarashi} and below) and one could expect a small contribution from high-order terms. Moreover, it is well known that $1/S$ series for staggered magnetization, transverse susceptibility, ground state energy and spin-wave stiffness of 2D AF calculated up to the third order in $1/S$ converge surprisingly fast even for $S\sim1$ despite the absence of a small parameter in the theory. \cite{monous,1sstatics,1sstif,igar1,igar2,1shamer} As a result the quantitative agreement is very good between $1/S$ expansion, numerical results and experiments. It should be stressed that quantum renormalization of these quantities is considerable for $S\sim1$. For instance, quantum fluctuations reduce the staggered magnetization in spin-$\frac12$ 2D AF from its bare value of 0.5 to about 0.3. Meantime this renormalization is described quantitatively by the first few terms of $1/S$ series.

We present in the present paper results of the spectrum $\epsilon_{\bf k}^{(3)}$ calculation in the third order in $1/S$ and demonstrate that $1/S$ series converges very fast in the whole BZ except for the vicinity of the point ${\bf k}=(\pi,0)$ in the case of $S\sim1$. In particular, we show that absolute values of the third order corrections to the spectrum at ${\bf k}=(\pi,0)$ are approximately equal to and only 2.5 times smaller than the second order ones for $S=1/2$ and $S=1$, respectively. Thus, our results demonstrate that, unlike other quantities, quantum renormalization of the spectrum near ${\bf k}=(\pi,0)$ for $S\sim1$ is described by slowly converging $1/S$ series. We find that the excitation energy in spin-$\frac12$ 2D AF $\epsilon_{(\pi,0)}^{(3)}\approx2.3241(2)$ is $3.2\%$ smaller than $\epsilon_{(\pi/2,\pi/2)}^{(3)}\approx2.4007(2)$ that improves (but still does not make perfect) the agreement with the recent experiments and numerical results (see Fig.~\ref{spec}). We suggest that despite the slow convergence of $1/S$ series the overall renormalization of the spectrum for $S=1$ might be small due to very small values of high-order $1/S$ corrections.

The rest of this paper is organized as follows. We present basic transformation of the Hamiltonian and describe the technique in Sec.~\ref{tec}. Spectrum renormalization is discussed in Sec.~\ref{ren}. Sec.~\ref{con} contains our conclusion. Expressions for self-energy parts in the third order in $1/S$ are presented in an appendix.

\section{Basic transformations and technique}
\label{tec}

The Hamiltonian of the Heisenberg AF on a square lattice with interaction between only nearest neighbor spins has the form
\begin{equation}
\label{ham0}
{\cal H} = \frac J2 \sum_{\langle i,j\rangle} {\bf S}_i  {\bf S}_j.
\end{equation}
We put exchange constant $J=1$ in all particular numerical calculations performed in the present paper. It is convenient to represent spins components in the local coordinate frame using Dyson-Maleev transformation in the following way:
\begin{eqnarray}
{\bf S}_j &=& S_j^x\hat x + (S_j^y\hat y + S_j^z\hat z)e^{i{\bf k}_0{\bf R}_j},\\
S^x_j &=& \sqrt{\frac S2} \left( a_j + a^\dagger_j - \frac{a_j^\dagger a_j^2}{2S} \right), \nonumber\\
S^y_j &=& -i\sqrt{\frac S2} \left( a_j - a^\dagger_j - \frac{a_j^\dagger a_j^2}{2S} \right), \label{md}\\
S^z_j &=& S - a_j^\dagger a_j,\nonumber
\end{eqnarray}
where $\hat x$, $\hat y$ and $\hat z$ are unit vectors along corresponding axes and ${\bf k}_0=(\pi,\pi)$ is AF vector. As a result one finds that the Hamiltonian (\ref{ham0}) acquires the form ${\cal H} = E_0 + \sum_{m=1}^6 {\cal H}_m$, where $E_0$ is the classical value of the ground state energy and ${\cal H}_m$ denote terms containing products of $m$ operators $a$ and $a^\dagger$. ${\cal H}_m=0$ for odd $m$ and one has for even $m$
\begin{eqnarray}
\label{h2}
{\cal H}_2 &=& \sum_{\bf k} 
\left[
E_{\bf k} a^\dagger_{\bf k}a_{\bf k} 
+ 
\frac{B_{\bf k}}{2} \left(a_{\bf k}a_{-\bf k} + a^\dagger_{\bf k}a^\dagger_{-\bf k}\right)
\right],\\
\label{h4}
{\cal H}_4 &=& -\frac{1}{2N}\sum_{{\bf k}_{1,2,3,4}}
a^\dagger_{-1}(J_{2+3}a^\dagger_{-2} + J_3a_2)a_3a_4,\\
\label{h6}
{\cal H}_6 &=& \frac{1}{8SN^2}\sum_{{\bf k}_{1,2,3,4,5,6}} 
J_{1+3+4}a^\dagger_{-1}a^\dagger_{-2}a_3a_4a_5a_6,
\end{eqnarray}
where $J_{\bf k} = 2(\cos k_x+\cos k_z)$, $E_{\bf k}=SJ_{\bf 0}$, $B_{\bf k}=SJ_{\bf k}$, $N$ is the number of spins in the lattice, we drop index $\bf k$ in Eqs.~(\ref{h4}) and (\ref{h6}) and the momentum conservation laws $\sum_{i=1}^4 {\bf k}_i = {\bf 0}$ and $\sum_{i=1}^6 {\bf k}_i = {\bf 0}$ are implied in Eqs.~(\ref{h4}) and (\ref{h6}), respectively. 

\begin{figure}
\centering
\includegraphics[scale=0.6]{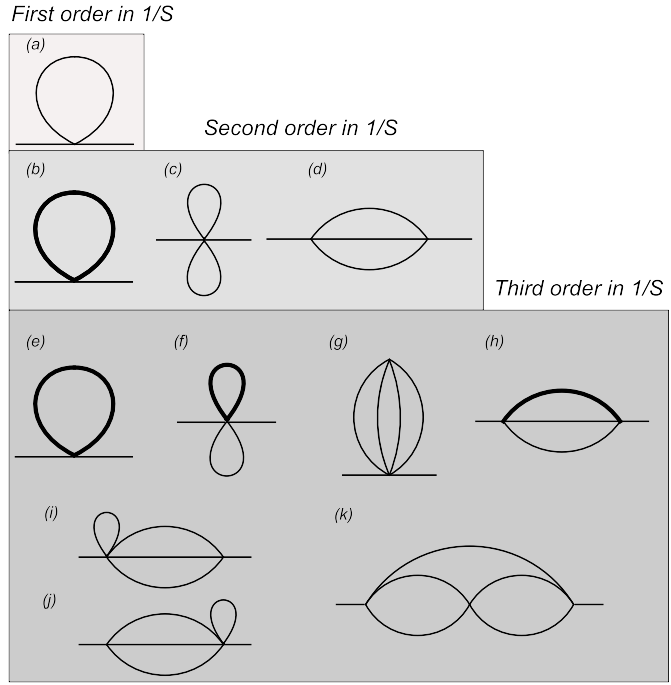}
\caption{Diagrams contributing to self-energy parts in first three orders in $1/S$. Bold lines in diagrams (b), (f) and (h) denote Green's functions of the first order in $1/S$ (i.e., Green's functions given by Eqs.~\eqref{gf} with self-energy parts calculated in the first order in $1/S$). The bold line in diagram (e) denotes Green's functions of the second order in $1/S$. Only diagrams (d) and (h)--(k) lead to the spectrum dispersion along the magnetic Brillouin zone boundary.
\label{diagrams}} 
\end{figure}

Introducing Green's functions $G(k) = \langle a_{\bf k}, a^\dagger_{\bf k} \rangle_\omega$, $F(k) = \langle a_{\bf k}, a_{-\bf k} \rangle_\omega$, ${\overline G}(k) = \langle a^\dagger_{-\bf k}, a_{-\bf k} \rangle_\omega$ and $F^\dagger (k) = \langle a^\dagger_{-\bf k}, a^\dagger_{\bf k} \rangle_\omega$, where $k=(\omega,{\bf k})$, we have two sets of Dyson equations for them one of which has the following form:
\begin{equation}
\label{eqfunc}
\begin{array}{l}
G(k) = G^{(0)}(k) + G^{(0)}(k){\overline \Sigma}(k)G(k) + G^{(0)}(k) [B_{\bf k} + \Pi(k)] F^\dagger(k),\\
F^\dagger(k) = {\overline G}^{(0)}(k) \Sigma(k)F^\dagger(k) + {\overline G}^{(0)}(k) [B_{\bf k} + \Pi^\dagger(k) ]G(k),
\end{array}
\end{equation}
where $G^{(0)}(k) = (\omega - E_{\bf k})^{-1}$ is the bare Green's function and $\Sigma(k)$, $\overline \Sigma(k)$, $\Pi(k)$ and $\Pi^\dagger(k)$ are self-energy parts. One obtains solving Eqs.~(\ref{eqfunc}) and similar set of equations for $\overline G(k)$ and $F(k)$
\begin{eqnarray}
G(k) &=& \frac{\omega + E_{\bf k} + \Sigma(k)}{{\cal D}(k)},\nonumber\\
\overline G(k) &=& \frac{-\omega + E_{\bf k} + \overline \Sigma(k)}{{\cal D}(k)},\nonumber\\
\label{gf}
F(k) &=& -\frac{B_{\bf k} + \Pi(k)}{{\cal D}(k)},\\
F^\dagger(k) &=& -\frac{B_{\bf k} + \Pi^\dagger(k)}{{\cal D}(k)},\nonumber
\end{eqnarray}
where
\begin{eqnarray}
\label{d}
{\cal D}(k) &=& \omega^2 - \left(\epsilon_{\bf k}^{(0)}\right)^2 - \Omega(k),\\
\label{spec0}
\epsilon_{\bf k}^{(0)} &=& \sqrt{E_{\bf k}^2 - B_{\bf k}^2} = S\sqrt{J_{\bf 0}^2-J_{\bf k}^2},\\
\label{o}
\Omega(k) &=& E_{\bf k}(\Sigma + \overline{\Sigma}) - B_{\bf k}(\Pi + \Pi^\dagger) - \omega (\Sigma - \overline{\Sigma}) - \Pi\Pi^\dagger + \Sigma \overline{\Sigma},
\end{eqnarray}
$G(k)=\overline G(-k)$, $\Sigma(k)=\overline \Sigma(-k)$ and $\epsilon_{\bf k}^{(0)}$ is the spin-wave spectrum in the linear spin-wave approximation (classical spectrum). Quantity $\Omega(k)$ given by Eq.~(\ref{o}) describes renormalization of the spin-wave spectrum square. We find $\Omega(k)$ within the first three orders in $1/S$ in the next section calculating corresponding diagrams for self-energy parts shown in Fig.~\ref{diagrams}.

It should be noted that we  do not use the conventional Bogolyubov transformation in the technique described to diagonalize the bilinear part of the Hamiltonian \eqref{h2}. As a result anomalous Green's functions $F(k)$ and $F^\dagger(k)$ arise and momenta lie in the chemical BZ that is twice as large as the magnetic one. Such an approach proved to be more convenient as intermediate calculations turn out to be more compact while the final results are equivalent to those obtained using the conventional approach. \cite{petmal,malanys,ifield,jpn1,jpn2}

\section{Spectrum renormalization}
\label{ren}

Although the spectrum renormalization within the first two orders in $1/S$ is well-known, we present here the corresponding expressions for the sake of completeness. 

\subsection{First order in $1/S$}

Only one diagram of the Hartree-Fock type shown in Fig.~\ref{diagrams}(a) contributes to the spectrum renormalization in the first order in $1/S$. The result can be represented in the form
\begin{eqnarray}
\label{sigmaa}
	\Sigma^{(a)}(k) &=& J_{\bf 0}(A+B),\\
\label{pia}
	\Pi^{(a)}(k) &=& J_{\bf k}A,\\
\label{pipa}
	\Pi^{\dagger(a)}(k) &=& J_{\bf k}(A+2B),\\
\label{spec1}
\epsilon_{\bf k}^{(1)} &=& \epsilon_{\bf k}^{(0)} \left( 1 + \frac{2(A+B)}{2S} \right) = \epsilon_{\bf k}^{(0)} \left( 1 + \frac{0.158}{2S} \right),
\end{eqnarray}
where the following two constants are introduced:
\begin{eqnarray}
	\label{a}
	A&=& \frac1N \sum_{\bf k} \frac{SJ_{\bf k}^2}{2J_0\epsilon^{(0)}_{\bf k}}\approx 0.2756,\\
	\label{b}
	B &=& -\frac1N \sum_{\bf k} \frac{SJ_{\bf 0} - \epsilon^{(0)}_{\bf k}}{2\epsilon^{(0)}_{\bf k}}\approx-0.1966.
\end{eqnarray}
It is seen from Eq.~\eqref{spec1} that renormalized spectrum remains flat on the BZ boundary in the first order in $1/S$ because $J_{\bf k}=0$ for $|k_x|=\pi-|k_z|$. We draw $\epsilon_{\bf k}^{(1)}$ for $S=1/2$ in Fig.~\ref{spec} using Eq.~\eqref{spec1}.

\subsection{Second order in $1/S$}

Diagrams (b)--(d) shown in Fig.~\ref{diagrams} contribute to self-energy parts in the second order in $1/S$. Diagram (b) is a schematic representation of the correction from diagram (a) of the second order in $1/S$ which arises after calculation of the diagram (a) with Green's functions of the first order in $1/S$ (i.e., Green's functions given by Eqs.~\eqref{gf} with self-energy parts given by Eqs.~\eqref{sigmaa}--\eqref{pipa}).

\subsubsection{Diagrams (b) and (c)}

It is convenient to group expressions for diagrams of the Hartree-Fock type (b) and (c) with the result
\begin{eqnarray}
\label{sigmabc}
	\Sigma^{(bc)}(k) &=& 0,\\
\label{pibc}
	\Pi^{(bc)}(k) &=& J_{\bf k} \frac{A(A-2B)}{2S},\\
\label{pipbc}
	\Pi^{\dagger(bc)}(k) &=& J_{\bf k} \frac{A^2+B^2+AB}{S},
\end{eqnarray}
where $A$ and $B$ are given by Eqs.~\eqref{a} and \eqref{b}, respectively. As $\Pi^{(bc)}(k),\Pi^{\dagger(bc)}(k) \propto J_{\bf k}$, these diagrams do not contribute to the spectrum dispersion along BZ boundary. 

\subsubsection{Diagram (d)}
\label{dd}

One obtains for corrections to self-energy parts from the diagram (d)
\begin{eqnarray}
\label{sigmad}
	\Sigma^{(d)}(k) &=& \frac{1}{N^2}\sum_{{\bf k}_1+{\bf k}_2+{\bf k}_3={\bf k}}
	\frac{1}{4 \epsilon _1 \epsilon _2 \epsilon _3 \left(\omega ^2-\left(\epsilon _1+\epsilon _2+\epsilon _3\right)^2\right)} \Bigl(
	\left(S J_0 \left(\epsilon _1+\epsilon _2+\epsilon _3\right)-\omega  \epsilon _1\right)
	\nonumber\\
	&&{}\times
	\left(\frac{1}{2} S^2J_{\bf k} J_1 J_2 J_3+S^2 J_2^2 J_3^2-2 S^2 J_0 J_{1-{\bf k}} J_2 J_3
		\right.\nonumber\\
	&&{}\left. +S^2 J_2 J_{2-{\bf k}} J_3 J_{3-{\bf k}}+J_{1-{\bf k}}^2 \left(S^2 J_0^2-\epsilon _2 \epsilon _3\right)\right)
		\nonumber\\
	&&{}
	-(\epsilon _1+\epsilon _2+\epsilon _3) \left(S^3 J_1 J_2^2 J_3 J_{2-{\bf k}}+S J_{\bf k} J_1 J_{1-{\bf k}} \left(S^2 J_0^2-\epsilon _2 \epsilon _3\right)\right)
	\Bigr),\\
\label{pid}
	\Pi^{(d)}(k) &=& \frac{1}{N^2}\sum_{{\bf k}_1+{\bf k}_2+{\bf k}_3={\bf k}}
	\frac{\epsilon _1+\epsilon _2+\epsilon _3}{4 \epsilon _1 \epsilon _2 \epsilon _3 \left(\omega ^2-\left(\epsilon _1+\epsilon _2+\epsilon _3\right)^2\right)} 	
	\nonumber\\
	&&{}\times
\left(-\frac{1}{2} S^3 J_1^3 J_2 J_3+2 S^3 J_0 J_2 J_{2-{\bf k}} J_3^2 -S^3 J_1 J_{2-{\bf k}}^2 J_2 J_3\right.
		\nonumber\\
	&&{}
	\left.
-S J_1 J_{2-{\bf k}} J_{3-{\bf k}} \left(S^2 J_0^2-\epsilon _2 \epsilon _3\right)\right),\\
\label{pipd}
	\Pi^{\dagger(d)}(k) &=& \frac{1}{N^2}\sum_{{\bf k}_1+{\bf k}_2+{\bf k}_3={\bf k}}
	\frac{\epsilon _1+\epsilon _2+\epsilon _3}{4 \epsilon _1 \epsilon _2 \epsilon _3 \left(\omega ^2-\left(\epsilon _1+\epsilon _2+\epsilon _3\right)^2\right)} 	
	\nonumber\\
	&&{}\times
	\left(-\frac{1}{2} S^3J_{\bf k}^2 J_1 J_2 J_3+2 S^3 J_0 J_{\bf k} J_1 J_{2-{\bf k}} J_3 -S^3 J_1^3 J_2 J_3 \right.
			\nonumber\\
	&&{}
	+2 S^3 J_0 J_1 J_{1-{\bf k}} J_2^2-S^3 J_{1-{\bf k}}^2 J_1 J_2 J_3
\nonumber\\
	&&{}\left.
-\left(2 S J_{\bf k} J_1^2+S J_1 J_2 J_3+2 S J_0 J_1 J_{1-{\bf k}}+S J_1 J_{2-{\bf k}} J_{3-{\bf k}}\right) \left(S^2 J_0^2-\epsilon _2 \epsilon _3\right)\right),
\end{eqnarray}
where we drop superscript $(0)$ in $\epsilon^{(0)}_{1,2,3}$ to light notations. Sums in Eqs.~\eqref{sigmad}--\eqref{pipd} over each momentum were calculated numerically by summing up $L^2$ points in BZ with some particular values of $L$ ranging from 20 to 200 and extrapolating the results to $L=\infty$ using the formula $A_\infty+A_1/L+A_2/L^2+\dots$, as it is done in previous papers. \cite{1shamer,igarashi,igar2} Appropriate symmetry of the summands was also used.

In accordance with previous results \cite{igarashi,igar2} we obtain that this diagram leads to a very small difference of 1.4\% between $\epsilon_{(\pi,0)}^{(2)}\approx2.35858$ and $\epsilon_{(\pi/2,\pi/2)}^{(2)}\approx2.39199$. It is seen from Table~\ref{points} that the second order corrections are much smaller than the first order ones in the whole BZ for all $S$. The spectrum $\epsilon_{\bf k}^{(2)}$ is presented in Fig.~\ref{spec} for $S=1/2$.

\begin{table}
\caption{Expressions are presented of the spin-wave spectrum $\epsilon_{\bf k}^{(3)}$ within the third order in $1/S$ in some representative points. Here $\epsilon_{\bf k}^{(0)}$ is the classical spectrum given by Eq.~\eqref{spec0}. The corresponding values of $\epsilon_{\bf k}^{(3)}$ are also shown for $S=1/2$. Notice the smallness of the second order $1/S$-corrections as compared with the first order ones for all points and all $S$. In contrast, the absolute value of the third order correction is approximately equal to the second oder one at ${\bf k}=(\pi,0)$ for $S=1/2$.
\label{points}
}
\begin{tabular}{|c|c|l|}
		\hline
$\displaystyle \bf k$	& \multicolumn{2}{c|}{$\displaystyle\epsilon_{\bf k}^{(3)}$}\\
&\multicolumn{1}{c}{arbitrary $S$} & \multicolumn{1}{c|}{$S=1/2$} \\
		\hline
		$\displaystyle\left(\frac{\pi}{4},0\right)$ & $\displaystyle\epsilon^{(0)}_{\bf k}\left(1+\frac{0.15795}{2 S}+\frac{0.02476}{(2 S)^2}-\frac{0.0033(3)}{(2 S)^3}\right)$ & 1.2290(3)\\
		$\displaystyle\left(\frac{\pi}{2},0\right)$ & $\displaystyle\epsilon^{(0)}_{\bf k}\left(1+\frac{0.15795}{2 S}+\frac{0.02879}{(2 S)^2}-\frac{0.0042(1)}{(2 S)^3}\right)$ & 2.0482(2)\\
		$\displaystyle\left(\frac{3\pi}{4},0\right)$ & $\displaystyle\epsilon^{(0)}_{\bf k}\left(1+\frac{0.15795}{2 S}+\frac{0.02538}{(2 S)^2}-\frac{0.0118(1)}{(2 S)^3}\right)$ & 2.3179(2)\\
		$\displaystyle\left(\pi,0\right)$ & $\displaystyle\epsilon^{(0)}_{\bf k}\left(1+\frac{0.15795}{2 S}+\frac{0.02134}{(2 S)^2}-\frac{0.0172(1)}{(2 S)^3}\right)$ & 2.3241(2)\\
		$\displaystyle\left(\frac{3\pi}{4},\frac{\pi}{4}\right)$ & $\displaystyle\epsilon^{(0)}_{\bf k}\left(1+\frac{0.15795}{2 S}+\frac{0.02967}{(2 S)^2}-\frac{0.0065(1)}{(2 S)^3}\right)$ & 2.3622(2)\\
		$\displaystyle\left(\frac{\pi}{2},\frac{\pi}{2}\right)$ & $\displaystyle\epsilon^{(0)}_{\bf k}\left(1+\frac{0.15795}{2 S}+\frac{0.03805}{(2 S)^2}+\frac{0.0043(1)}{(2 S)^3}\right)$ & 2.4007(2)\\
		$\displaystyle\left(\frac{3\pi}{4},\frac{3\pi}{4}\right)$ & $\displaystyle\epsilon^{(0)}_{\bf k}\left(1+\frac{0.15795}{2 S}+\frac{0.02914}{(2 S)^2}-\frac{0.0005(1)}{(2 S)^3}\right)$ & 1.6781(2)\\
		\hline
\end{tabular}
\end{table}

\subsection{Third order in $1/S$}

One has to analyze in this order diagrams shown in Fig.~\ref{diagrams}(e)--(k). Diagram (e) represents the second order correction from the diagram (a) which should be calculated using Eqs.~\eqref{sigmaa}--\eqref{pipd}. Bold lines in diagrams (f) and (h) denote Green's functions of the first order in $1/S$. Expressions for self-energy parts in this order are quite complicated and the reader is referred to \ref{third} for some detail of their calculation. It can be shown (see Appendix~\ref{third}) that diagrams of the Hartree-Fock type presented in Fig.~\ref{diagrams}(e)--(g) do not change along BZ boundary so that only diagrams (h)--(k) give rise to the spectrum dispersion in these directions in this order. In contrast to the second order corrections \eqref{sigmad}--\eqref{pipd} one has to calculate triple sums over momenta in the third order. This procedure requires pretty much computer time. Then, we focus on short-wavelength magnons as their spectrum renormalization is expected to be most pronounced and calculate $\epsilon_{\bf k}^{(3)}$ in a number of points with $|{\bf k}|,|{\bf k}-{\bf k}_0|\ge\pi/8$. The results are presented in Fig.~\ref{spec} (for $S=1/2$) and in Table~\ref{points}. 

It is seen from Table~\ref{points} that the third order corrections are noticeable only for $S\sim1$ and only in the vicinity of the point ${\bf k}=(\pi,0)$. In particular, absolute values of the third and the second order corrections in $1/S$ are approximately equal to each other at ${\bf k}=(\pi,0)$ for $S=1/2$. The excitation energy in spin-$\frac12$ 2D AF $\epsilon_{(\pi,0)}^{(3)}\approx2.3241(2)$ is $3.2\%$ smaller than $\epsilon_{(\pi/2,\pi/2)}^{(3)}\approx2.4007(2)$ that improves the agreement with the recent experiments and numerical results leaving it, however, far from being perfect. Thus, our calculations demonstarte that quantum renormalization of the spectrum near ${\bf k}=(\pi,0)$ for $S=1/2$ is described by slowly converging $1/S$ series. 

It is also seen from Table~\ref{points} that at ${\bf k}=(\pi,0)$ the third order correction is only $2.5$ times smaller than the second order one for $S=1$. Thus, one can expect slow convergence of $1/S$ series near ${\bf k}=(\pi,0)$ also for $S=1$. Meantime the overall renormalization of the spectrum would be small due to very small values of high-order $1/S$ terms.

\section{Conclusion}
\label{con}

To conclude, we calculate the spin-wave spectrum of 2D AF on a square lattice in the third order in $1/S$ to examine the convergence of $1/S$ series. Within the first two orders we recover the previous results \cite{igarashi,igar2} showing that the second-order corrections are much smaller than the first order ones in the whole BZ and for all $S$ (see Table~\ref{points}). Our calculation of the spectrum in the next order demonstrates that the third order corrections to the spectrum are much smaller than the second order ones in the whole BZ except for the vicinity of the point ${\bf k}=(\pi,0)$ in the case of $S\sim1$. In particular, their absolute values are approximately equal at ${\bf k}=(\pi,0)$ for $S=1/2$ (see Table~\ref{points} and Fig.~\ref{spec}). Thus, our results demonstrate that, unlike other quantities, quantum renormalization of the spectrum near ${\bf k}=(\pi,0)$ for $S\sim1$ is described by slowly converging $1/S$ series. We find that third order corrections for the spectrum improves the agreement with the recent experiments and numerical results in spin-$\frac12$ 2D AF. We expect slow convergence of $1/S$ series near ${\bf k}=(\pi,0)$ also for $S=1$ while the overall renormalization of the spectrum would be small in this case due to very small values of high-order $1/S$ corrections.

\begin{acknowledgments}

This work was supported by President of Russian Federation (grant MK-329.2010.2), RFBR grants 09-02-00229, and Russian Programs "Quantum Macrophysics", "Strongly correlated electrons in semiconductors, metals, superconductors and magnetic materials" and "Neutron Research of Solids".

\end{acknowledgments}

\appendix

\section{Expressions for the third-order diagrams}
\label{third}

We present in this Appendix expressions for self-energy parts in the third order in $1/S$ which originate from diagrams shown in Fig.~\ref{diagrams}(e)--(k). Simple codes have been written in Mathematica software to generate the majority of these expressions. To make it compact we present below expression for the sum of anomalous self-energy parts $\Pi(k) + \Pi^\dagger(k)$ rather than for $\Pi(k)$ and $\Pi^\dagger(k)$ separately because only this sum contributes to the spectrum renormalization in this order (see Eq.~\eqref{o}).

\subsection{Diagrams (e) and (f)}

It is convenient to group contributions from diagrams of the Hartree-Fock type shown in Fig.~\ref{diagrams}(e) and (f). One has after simple calculations using expressions \eqref{sigmaa}--\eqref{pipd}
\begin{eqnarray}
\label{sigmaef}
&&\Sigma^{(ef)}(k) = -J_0 \frac{A^3}{(2S)^2} + \frac{1}{N^3}\sum_{{\bf k}_1+{\bf k}_2+{\bf k}_3+{\bf k}_4=0}\frac{S}{32 \epsilon _1 \epsilon _2 \epsilon _3 \epsilon _4 \left(\epsilon _1+\epsilon _2+\epsilon _3+\epsilon _4\right)^2}
\nonumber\\
&&{}\times
\Bigl(
-3 S^2 J_1^3 J_2 J_3 J_4 \epsilon _1+J_1 J_4 \left(-15 S^2 J_2^3 J_3 \epsilon _1+8 S^2 J_0 J_2^2 J_{1+4} \epsilon _4
\right.
\nonumber\\
&&{}\left.
+8 S^2 J_0 J_3 J_4 J_{2+4} \left(\epsilon _1+\epsilon _4\right)
+2 J_2 J_3 \left(3 \epsilon _1 \epsilon _2 \epsilon _3-S^2 \left(3 J_0^2+2 J_{1+4}^2\right) \epsilon _4\right)\right)
\nonumber\\
&&{}+4 \left(J_4^2 J_{1+4}^2 \epsilon _1 \left(-S^2 J_0^2+\epsilon _2 \epsilon _3\right)-3 J_2^2 J_4^2 \left(S^2 J_3^2 \epsilon _1+\epsilon _2 \left(S^2 J_0^2-\epsilon _1 \epsilon _3\right)\right)
\right.
\nonumber\\
&&{}
+J_2 \left(4 S^2 J_0^3 J_4 J_{2+4} \epsilon _2+2 J_0 J_4 \epsilon _1 \left(S^2 J_3 \left(2 J_4 J_{1+4}+J_3 J_{2+4}\right)-2 J_{2+4} \epsilon _2 \epsilon _3\right)
\right.\nonumber\\
&&{}+S^2 J_0^2 J_{3+4} \left(J_3 J_{2+4} \left(\epsilon _1-\epsilon _4\right)
-J_4 J_{1+4} \epsilon _4\right)
\nonumber\\
&&{}\left.\left.
+J_4 J_{3+4} \epsilon _1 \left(-S^2 J_3 J_4 J_{2+4}+J_{1+4} \epsilon _3 \epsilon _4\right)\right)\right)
\Bigr),\\
&&
\label{pief}
\Pi^{(ef)}(k) + \Pi^{\dagger(ef)}(k) = J_{\bf k} \frac{A(B^2+(B-A)^2)}{2S^2} + 2\frac{J_{\bf k}}{J_0}\Sigma^{(ef)}(k) 
\nonumber\\
&&
+ 
\frac{J_{\bf k}}{J_0}\frac{1}{8N^3}\sum_{{\bf k}_1+{\bf k}_2+{\bf k}_3+{\bf k}_4=0}\frac{1}{ \epsilon _1 \epsilon _2 \epsilon _3 \epsilon _4 \left(\epsilon _1+\epsilon _2+\epsilon _3+\epsilon _4\right)}
\nonumber\\
&&{}\times
\Bigl(S J_3 J_4 \left(S^2 J_1^3 J_2-2 S^2 J_0 J_1 J_{2+4} J_4 + \left(J_1 J_2-2 J_0 J_{3+4}+2 J_3 J_4\right) \left(S^2 J_0^2-\epsilon _1 \epsilon _2\right)\right)\Bigr),
\end{eqnarray}
where $A$ and $B$ are given by Eqs.~\eqref{a} and \eqref{b}, respectively. Sums in Eqs.~\eqref{sigmaef} and \eqref{pief} were calculated as it is described above in Sec.~\ref{dd} for diagram (d) with $20\le L\le96$. These sums arise after taking into account in Green's functions involving in the diagram (e) contributions to self-energy parts from the diagram (d). We obtain numerically from Eqs.~\eqref{sigmaef} and \eqref{pief} $\Sigma^{(ef)}(k) = \frac{0.0598(2)}{(2S)^2}$ and $\Pi^{(ef)}(k) + \Pi^{\dagger(ef)}(k) = J_{\bf k}\frac{0.1446(1)}{(2S)^2}$.

\subsection{Diagram (g)}

Corrections to self-energy parts from another Hartree-Fock diagram shown in Fig.~\ref{diagrams}(g) have the form
\begin{eqnarray}
\label{sigmag}
&&\Sigma^{(g)}(k) = \frac{1}{16N^3}\sum_{{\bf k}_1+{\bf k}_2+{\bf k}_3+{\bf k}_4=0}
\frac{1}{\epsilon _1 \epsilon _2 \epsilon _3 \epsilon _4 \left(\epsilon _1+\epsilon _2+\epsilon _3+\epsilon _4\right)}
\nonumber\\
&&\times{}
\Bigl(S J_3 J_4 \left(S^2 J_1^3 J_2-2 S^2 J_0 J_1 J_{2+4} J_4 + \left(J_1 J_2-2 J_0 J_{3+4}+2 J_3 J_4\right) \left(S^2 J_0^2-\epsilon _1 \epsilon _2\right)\right)\Bigr),\\
\label{pig}
&&\Pi^{(g)}(k) + \Pi^{\dagger(g)}(k) = \frac{J_{\bf k}}{J_0}\frac{1}{N^3}\sum_{{\bf k}_1+{\bf k}_2+{\bf k}_3+{\bf k}_4=0}
\frac{1}{8S\epsilon _1 \epsilon _2 \epsilon _3 \epsilon _4 \left(\epsilon _1+\epsilon _2+\epsilon _3+\epsilon _4\right)}
\nonumber\\
&&\times{}
J_{1+2} \Bigl(S^2 \left(-2 S^2 J_0^3 J_{1} J_{2}+S^2 J_0^4 J_{1+2}-3 S^2 J_0 J_{1}^2 J_{3} J_{4}+J_0^2 \left(S^2 J_{1+3} \left(2 J_{2} J_{3}+J_{1} J_{4}\right)
\right.\right.
\nonumber\\
&&\left.\left.
-2 J_{1+2} \epsilon _1 \epsilon _2\right)+J_{1} J_{4} \left(S^2 J_{2} J_{1+2} J_{3}+J_{1+3} \epsilon _2 \epsilon _3\right)\right)
\nonumber\\
&&{}+\left(J_{1+2} \epsilon _1 \epsilon _2 \epsilon _3+2 S^2 J_{2} \left(-J_{3} J_{1+3} \epsilon _1+J_0 J_{1} \epsilon _3\right)\right) \epsilon _4\Bigr).
\end{eqnarray}
Notice the identity of sums in Eqs.~\eqref{sigmag} and \eqref{pief}. Sums in Eqs.~\eqref{sigmag} and \eqref{pig} were calculated with $20\le L\le96$. We obtain numerically from Eqs.~\eqref{sigmag} and \eqref{pig} $\Sigma^{(g)}(k) = -\frac{0.05892(2)}{(2S)^2}$ and $\Pi^{(g)}(k) + \Pi^{\dagger(g)}(k) = -J_{\bf k}\frac{0.02800(4)}{(2S)^2}$.

\subsection{Diagram (h)}

It is convenient to divide corrections from diagram shown in Fig.~\ref{diagrams}(h) into two parts: $\Sigma^{(h)}(k) = \Sigma_1^{(h)}(k)+ \Sigma_2^{(h)}(k)$,  $\Pi^{(h)}(k) = \Pi_1^{(h)}(k)+ \Pi_2^{(h)}(k)$, and $\Pi^{\dagger(h)}(k) = \Pi_1^{\dagger(h)}(k)+ \Pi_2^{\dagger(h)}(k)$, where the first terms arise after taking into account first order $1/S$ corrections to self-energy parts in numerators of Green's functions in diagram (d). As a result one has for them
\begin{eqnarray}
\label{sigmah}
	&&\Sigma^{(h)}_1(k) = \frac{1}{N^2}\sum_{{\bf k}_1+{\bf k}_2+{\bf k}_3={\bf k}}
	\frac{1}{4 \epsilon _1 \epsilon _2 \epsilon _3 \left(\omega ^2-\left(\epsilon _1+\epsilon _2+\epsilon _3\right)^2\right)} \Bigl((\epsilon _1+\epsilon _2+\epsilon _3) 
	\nonumber\\
	&&{}\times
	\left(\frac12(3A+B) S^2J_0J_{\bf k} J_1 J_2 J_3 - (3A+2B) S^2J_0^2 J_{1-{\bf k}} (J_{\bf k} J_1 + 2J_2J_3) 
	\right.
		\nonumber\\
	&&{}
+ AJ_{\bf k} J_1 J_{1-{\bf k}}\epsilon _2 \epsilon _3 + (3A+B)S^2J_0 J_2^2 J_3^2 
	-(3A+2B) S^2 J_1^2 J_{1-{\bf k}} J_2 J_3 
		\nonumber\\
	&&{}\left.+ 3(A+B)S^2J_0 J_2 J_{2-{\bf k}} J_3 J_{3-{\bf k}} + 3(A+B) S^2 J_0^3J_{1-{\bf k}}^2 - (A+B) J_0J_{1-{\bf k}}^2\epsilon _2 \epsilon _3 \right)
		\nonumber\\
	&&{}
	- \omega  \epsilon _1 
	\left( 
	2ASJ_2^2 J_3^2 -
	(2A+B)2SJ_0 J_{1-{\bf k}}J_2J_3 
		\right.\nonumber\\
	&&{}+\left.
	2S (A+B)(J_2 J_{2-{\bf k}} J_3 J_{3-{\bf k}} + J_0^2J_{1-{\bf k}}^2) +
	A SJ_{\bf k} J_1 J_2 J_3
	\right)
	\Bigr),\\
\label{pih}
	&&\Pi^{(h)}_1(k) + \Pi^{\dagger(h)}_1(k) = 
\frac{1}{N^2}\sum_{{\bf k}_1+{\bf k}_2+{\bf k}_3={\bf k}}
	\frac{\epsilon _1+\epsilon _2+\epsilon _3}{4 \epsilon _1 \epsilon _2 \epsilon _3 \left(\omega ^2-\left(\epsilon _1+\epsilon _2+\epsilon _3\right)^2\right)} 	
		\nonumber\\
	&&{}\times		
		\Bigl(
		6 (A+B) S^2 J_0^3 J_2 J_{2-\bf k}
		-S^2 J_0^2 \left(J_3 \left((3 A+2 B) J_1 J_2+6 (A+B) J_{1-{\bf k}} J_{2-{\bf k}}\right)
		\right.\nonumber\\
	&&{}\left.+2 (3 A+2 B) J_1^2 J_{\bf k}\right)+\frac{1}{2} J_3 \left(-S^2 J_1 J_2 \left((9 A+4 B) J_1^2\right.\right.
				\nonumber\\
	&&{}\left.\left.
		+12 (A+B) J_{3-{\bf k}}^2+(3 A+2 B) J_{\bf k}^2\right)+2 \left(A J_1 J_2+2 (A+B) J_{1-{\bf k}} J_{2-{\bf k}}+2 A J_3 J_{\bf k}\right) \epsilon _1 \epsilon _2\right)
				\nonumber\\
	&&{}
	+2 J_0 \left(2 (3 A+2 B) S^2 J_1^2 J_2 J_{2-{\bf k}}+(A+B) J_3 \left(3 S^2 J_2 J_{1-{\bf k}} J_{\bf k}-J_{3-{\bf k}} \epsilon _1 \epsilon _2\right)\right)
		\Bigr).
\end{eqnarray}
Expressions for $\Sigma_2^{(h)}(k)$, $ \Pi_2^{(h)}(k)$, and $\Pi_2^{\dagger(h)}(k)$ can be easily obtained from Eqs.~\eqref{sigmad}, \eqref{pid} and \eqref{pipd} taking into account the first-order renormalization of the spectrum $\epsilon_{\bf k}^{(1)}=\epsilon_{\bf k}^{(0)}(1+(A+B)/S)$ and the fact that one has to put $\epsilon_{\bf k}^{(1)}$ instead of $\omega$ calculating the third-order correction to the spectrum. Sums in Eqs.~\eqref{sigmah} and \eqref{pih} were calculated with $20\le L\le200$.

\subsection{Diagrams (i) and (j)}

Grouping expressions for diagrams shown in Fig.~\ref{diagrams}(i) and (j) one obtains
\begin{eqnarray}
\label{sigmaij}
	&&\Sigma^{(ij)}(k) = \frac{1}{N^2}\sum_{{\bf k}_1+{\bf k}_2+{\bf k}_3={\bf k}}
	\frac{1}{4 \epsilon _1 \epsilon _2 \epsilon _3 \left(\omega ^2-\left(\epsilon _1+\epsilon _2+\epsilon _3\right)^2\right)} \Bigl(
	\left(S J_0 \left(\epsilon _1+\epsilon _2+\epsilon _3\right)-\omega  \epsilon _1\right)
	\nonumber\\
	&&{}\times
	\left(B SJ_{\bf k} J_1 J_2 J_3 + 2BS J_2^2 J_3^2 - 2(A+B) S J_0 J_{1-{\bf k}} J_2 J_3 + 2AS J_2 J_{2-{\bf k}} J_3 J_{3-{\bf k}} \right.
			\nonumber\\
	&&{}\left.
	+ \frac{2}{S}AJ_{1-{\bf k}}^2 \left(S^2 J_0^2-\epsilon _2 \epsilon _3\right)\right)
			+ 2AS J_2 J_{2-{\bf k}} J_3 J_{3-{\bf k}}\left(S J_0 \left(\epsilon _1+\epsilon _2+\epsilon _3\right)+\omega  \epsilon _1\right)\nonumber\\
	&&{}
	-(\epsilon _1+\epsilon _2+\epsilon _3) \left(\frac12(3A+2B)S^2 J_1 J_2^2 J_3 J_{2-{\bf k}} + (A+B) J_{\bf k} J_1 J_{1-{\bf k}} \left(S^2 J_0^2-\epsilon _2 \epsilon _3\right)\right)
	\Bigr),\\
\label{piij}
	&&\Pi^{(ij)}(k) + \Pi^{\dagger(ij)}(k) = 
\frac{1}{N^2}\sum_{{\bf k}_1+{\bf k}_2+{\bf k}_3={\bf k}}
	\frac{\epsilon _1+\epsilon _2+\epsilon _3}{4 \epsilon _1 \epsilon _2 \epsilon _3 \left(\omega ^2-\left(\epsilon _1+\epsilon _2+\epsilon _3\right)^2\right)} 	
		\Bigl(-B S^2J_{\bf k}^2 J_1 J_2 J_3
	\nonumber\\
	&&{}
 + (3A+2B)S^2 J_0 J_{\bf k} J_1 J_{2-{\bf k}} J_3 - 3BS^2 J_1^3 J_2 J_3 + 2(3A+2B) S^2 J_0 J_1 J_{1-{\bf k}} J_2^2
	\nonumber\\
	&&{}
 - 4AS^2 J_{1-{\bf k}}^2 J_1 J_2 J_3 - 4AS^2 J_0^2 J_{1-{\bf k}} J_{2-{\bf k}} J_3 - (4B J_{\bf k} J_1^2 + 2BJ_1 J_2 J_3 
 	\nonumber\\
	&&{}
- 2(A+B) J_0 J_1 J_{1-{\bf k}} + 2A J_1 J_{2-{\bf k}} J_{3-{\bf k}}) \left(S^2 J_0^2-\epsilon _2 \epsilon _3\right)\Bigr).
\end{eqnarray}
Sums in Eqs.~\eqref{sigmaij} and \eqref{piij} were calculated with $20\le L\le200$.

\subsection{Diagram (k)}

Expressions stemming from this diagram are very cumbersome. We present here only expression for $\Sigma^{(k)}(k)$ in the most compact form (i.e., before integration over energies) for the particular case of the momentum $\bf k$ lying on BZ boundary (i.e., for $|k_x|=\pi-|k_z|$ and $J_{\bf k}=0$) that has the form
\begin{eqnarray}
\label{sigmak}
&&\Sigma^{(k)}(k)= -i\frac{1}{N^3} \sum \nonumber\\
&&{}
\Bigl(-8 F_{1} G_{5} {\overline G}_{2} {\overline G}_{3} {\overline G}_{4} J_{1-{\bf k}} J_{2} J_{2-{\bf k}}
-8 F_{4} G_{1} G_{3} {\overline G}_{2} {\overline G}_{5} J_{1} J_{1-{\bf k}} J_{4-{\bf k}}
-8 F_{4} F_{5} G_{1} G_{3} {\overline G}_{2} J_{1} J_{4} J_{4-{\bf k}}\nonumber\\
&&{}
-8 F_{2} G_{3} G_{5} {\overline G}_{1} {\overline G}_{4} 
\left(2 J_{1-{\bf k}}^2 J_{2}+J_{2-{\bf k}} J_{4} J_{4-{\bf k}}\right)
-8 F_{1} F_{2} F_{5} G_{3} G_{4} J_{1} J_{1-{\bf k}} J_{2+4}\nonumber\\
&&{}
-8 F_{5} G_{2} G_{3} G_{4} {\overline G}_{1} J_{1-{\bf k}} J_{2} J_{2+4}-8 F_{3} {\overline G}_{1} {\overline G}_{2} {\overline G}_{4} {\overline G}_{5} J_{1-{\bf k}} J_{4} J_{2+4}
-8 F_{4} F_{5} G_{2} G_{3} {\overline G}_{1} J_{3} J_{4} J_{2+4}
\nonumber\\
&&{}
-8 F_{2} F_{3} {\overline G}_{1} {\overline G}_{4} {\overline G}_{5} J_{3} J_{4} J_{2+4} -8F_{1} {\overline G}_{2} {\overline G}_{3} {\overline G}_{4} {\overline G}_{5} J_{3-{\bf k}} J_{4} J_{2+4}
-8 F_{2} F_{4} F_{5} G_{1} G_{3} J_{1} J_{4-{\bf k}} J_{2+4}
\nonumber\\
&&{}
-8 F_{1} F_{4} F_{5} G_{2} G_{3} J_{2} J_{4-{\bf k}} J_{2+4}-8 G_{3} G_{5} {\overline G}_{1} {\overline G}_{2} {\overline G}_{4} J_{2-{\bf k}} J_{4-{\bf k}} J_{2+4}
\nonumber\\
&&{}
-8 G_{1} {\overline G}_{2} {\overline G}_{3} {\overline G}_{4} {\overline G}_{5} J_{2-{\bf k}} J_{4-{\bf k}} J_{2+4}
-8 F_{1} F_{4} F_{5} {\overline G}_{2} {\overline G}_{3} J_{2-{\bf k}} 
\left(J_{1-{\bf k}} J_{3}+J_{4} J_{2+4}\right)
\nonumber\\
&&{}
-8 F_{1} G_{3} G_{5} {\overline G}_{2} {\overline G}_{4} 
\left(J_{1} J_{1-{\bf k}}^2+J_{3} J_{4-{\bf k}} J_{2+4}\right)
-4 F_{1} F_{2} F_{3} F_{4} F_{5} \left(J_{1} 
\left(2 J_{1-{\bf k}}^2+3 J_{2}^2\right)
\right.\nonumber\\
&&{}
\left.+2 J_{4} \left(J_{1-{\bf k}} J_{4-{\bf k}}+2 J_{3-{\bf k}} J_{2+4}\right)\right)\nonumber\\
&&{}
-4 F_{2} F_{3} F_{4} F_{5} {\overline G}_{1} 
\left(4 J_{1-{\bf k}} J_{2} J_{3}+J_{2-{\bf k}} \left(J_{1} J_{3}+2 J_{4-{\bf k}} J_{2+4}\right)\right)
\nonumber\\
&&{}
-8 F_{2} F_{5} G_{3} G_{4} {\overline G}_{1} J_{1-{\bf k}} \left(2 J_{2} J_{4}+J_{1-{\bf k}} J_{3+4}\right)\nonumber\\
&&{}
-4 F_{4} F_{5} G_{3} {\overline G}_{1} {\overline G}_{2} \left(4 J_{1-{\bf k}} \left(J_{1-{\bf k}}^2+J_{4}^2\right)+J_{2-{\bf k}} \left(J_{1} J_{2}+4 J_{4-{\bf k}} J_{3+4}\right)\right)
\nonumber\\
&&{}
-8 F_{2} F_{3} F_{4} F_{5} G_{1} J_{4-{\bf k}} \left(J_{2-{\bf k}} J_{2+4}+J_{1} J_{5}\right)\nonumber\\
&&{}
-8 F_{2} F_{5} G_{3} {\overline G}_{1} {\overline G}_{4} J_{1-{\bf k}} \left(J_{2} J_{4}+J_{3} J_{5}\right)
-16 F_{1} F_{3} F_{5} G_{4} {\overline G}_{2} J_{1-{\bf k}} \left(J_{2-{\bf k}} J_{4}+J_{3-{\bf k}} J_{5}\right)\nonumber\\
&&{}
-8 F_{2} F_{5} {\overline G}_{1} {\overline G}_{3} {\overline G}_{4} \left(2 J_{1-{\bf k}} J_{2} J_{4}+J_{1-{\bf k}}^2 J_{3+4}+J_{1} J_{3-{\bf k}} J_{5}\right)\nonumber\\
&&{}
-8 F_{3} F_{4} F_{5} {\overline G}_{1} {\overline G}_{2} \left(2 J_{1-{\bf k}}^2 J_{2}+J_{2} J_{4}^2+J_{1-{\bf k}} J_{4} J_{3+4}+J_{4-{\bf k}} \left(J_{2-{\bf k}} J_{4}+J_{1} J_{3+4}\right)+J_{3} J_{4} J_{5}\right)\nonumber\\
&&{}
-8 F_{3} F_{4} F_{5} G_{2} {\overline G}_{1} \left(2 J_{1-{\bf k}}^2 J_{2}+J_{2} J_{4}^2+J_{1-{\bf k}} J_{4} J_{3+4}+\left(J_{3} J_{4}+J_{3-{\bf k}} J_{4-{\bf k}}\right) J_{5}\right)\nonumber\\
&&{}
-8 F_{1} F_{3} F_{5} {\overline G}_{2} {\overline G}_{4} \left(J_{1} \left(J_{2} J_{4}+J_{1-{\bf k}} J_{3+4}+2 J_{3} J_{5}\right)+2 J_{1-{\bf k}} \left(J_{2-{\bf k}} J_{4}+J_{3-{\bf k}} J_{5}\right)\right)\nonumber\\
&&{}
-8 F_{2} G_{5} {\overline G}_{1} {\overline G}_{3} {\overline G}_{4} \left(2 J_{1-{\bf k}}^2 J_{2}+J_{4-{\bf k}} \left(J_{1} J_{3+4}+J_{3-{\bf k}} J_{5}\right)\right)\nonumber\\
&&{}
-8 F_{1} F_{2} F_{4} F_{5} {\overline G}_{3} (2 J_{1-{\bf k}}^2 J_{2-{\bf k}}+J_{1} \left(2 J_{1-{\bf k}} J_{2}+J_{4} J_{3+4}\right)
\nonumber\\
&&{}
+J_{4} \left(2 J_{2-{\bf k}} J_{4}+J_{2} J_{4-{\bf k}}+2 J_{3-{\bf k}} J_{5}\right))\nonumber\\
&&{}
-4 F_{1} F_{4} F_{5} G_{3} {\overline G}_{2} \left(2 J_{2-{\bf k}} J_{4} J_{2+4}+2 J_{4-{\bf k}} \left(J_{3} J_{3+4}+J_{1-{\bf k}} J_{5}\right)+J_{1} \left(J_{2} J_{3}+2 J_{4} J_{5}\right)\right)\nonumber\\
&&{}
-8F_{2} F_{5} G_{1} {\overline G}_{3} {\overline G}_{4} \left(\left(J_{1} J_{2}+2 J_{1-{\bf k}} J_{2-{\bf k}}\right) J_{4-{\bf k}} + J_{1} J_{3} J_{5-{\bf k}}\right)
\nonumber\\
&&{}
-8 F_{1} F_{5} G_{3} G_{4} {\overline G}_{2} \left(J_{1} J_{1-{\bf k}} J_{4}+J_{2} J_{3} J_{5-{\bf k}}\right)\nonumber\\
&&{}
-8 F_{2} F_{3} F_{5} G_{1} {\overline G}_{4} \left(J_{1} J_{1-{\bf k}} J_{4-{\bf k}}+J_{2-{\bf k}} \left(J_{2} J_{4-{\bf k}}+J_{3} J_{5-{\bf k}}\right)\right)
\nonumber\\
&&{}
-8 F_{4} G_{1} {\overline G}_{2} {\overline G}_{3} {\overline G}_{5} J_{2-{\bf k}} \left(J_{2} J_{4-{\bf k}}+J_{3} J_{5-{\bf k}}+J_{1} J_{2+5}\right)\nonumber\\
&&{}
-8F_{1} F_{5} {\overline G}_{2} {\overline G}_{3} {\overline G}_{4} \left(J_{2-{\bf k}} \left(J_{2} J_{4}+J_{3} J_{5}\right) + \left(J_{1} J_{2}+2 J_{1-{\bf k}} J_{2-{\bf k}}\right) J_{2+5}\right)\nonumber\\
&&{}
-8 F_{1} F_{2} F_{3} F_{4} G_{5} \left(J_{2} J_{2-{\bf k}} J_{4}+J_{2}^2 J_{4-{\bf k}}+J_{2-{\bf k}} \left(J_{3} J_{5}+2 J_{1-{\bf k}} J_{2+5}\right)+J_{1} \left(J_{1-{\bf k}} J_{4}+J_{2} J_{2+5}\right)\right)\nonumber\\
&&{}
-8 F_{1} F_{3} G_{5} {\overline G}_{2} {\overline G}_{4} \left(2 J_{1-{\bf k}}^2 J_{2-{\bf k}}+J_{3} J_{4-{\bf k}} J_{5}+J_{1} \left(2 J_{1-{\bf k}} J_{2}+J_{5} J_{3+5}\right)\right)\Bigr),
\end{eqnarray}
where the momentum conservation laws $k_1+k_2+k_3=k_1+k_4+k_5=k$ are implied. The expression for $\Sigma^{(k)}(k)$ for arbitrary momentum and those for anomalous self-energy parts are much more cumbersome than Eq.~\eqref{sigmak} and we do not present them here. Corresponding sums over momenta were calculated with $16\le L\le64$.

\section*{References}

\bibliography{3order} 

\end{document}